# On the alternative description of complex holomorphic and Lorentz geometries in four dimensions
## II. Appendix

S.I.Tertychniy *

Institute for Theoretical Physics
University
D-50923 Cologne
Germany

**Abstract**

This supplementary part of the paper gr-qc 9312038 contains the necessary proofs of the claims stated in the main part.

**Appendix. The proofs.**

*Proof of the Lemma* **1.2**  Let $\alpha = \mu + \nu$, $\mu \in {}^{\pm}\Lambda^2$, $\nu \in {}^{\mp}\Lambda^2$. We may assume $\alpha \neq 0$ otherwise there is nothing to prove. For every $\beta \in {}^{\pm}\Lambda^2$ $\alpha \wedge \beta = \mu \wedge \beta$ (since $\nu \wedge \beta \in {}^{\mp}\Lambda^2 \wedge {}^{\pm}\Lambda^2 = 0$). Thus $\mu \wedge \beta = 0$. But $\mu \wedge \gamma = 0$ for every $\gamma \in {}^{\mp}\Lambda^2$ as well. Finally, every 2-form can be represented as a sum of the form $\beta + \gamma$. Hence $\mu \wedge \Lambda^2 = 0$ and therefore $\mu = 0$.  Q.E.D.

*Proof of the Lemma* **1.5**  Let $\alpha \neq 0$ be a lobe element such that $\alpha \wedge \alpha \neq 0$. Choose any other nonzero lobe element $\beta$ independent on $\alpha$. We may assume $\beta \wedge \beta \neq 0$ otherwise there is nothing to prove. Let $\gamma = \alpha + k\beta$ for some complex number $k$. Due to the linear independence of $\alpha$ and $\beta$ $\gamma \neq 0$ for every $k$. But since $\gamma \wedge \gamma = \alpha \wedge \alpha + 2k\alpha \wedge \beta + k^2 \beta \wedge \beta$ and $\Lambda^4$ is 1-dimensional there exist such $k$ that $\gamma \wedge \gamma = 0$.  Q.E.D.

*Proof of the Lemma* **1.7**  Let $\alpha$ be a nonzero simple lobe element. Let $\beta$ be another lobe element such that $\alpha \wedge \beta \neq 0$; it exists due to the corollary **1.3** and is obviously independent with $\alpha$. Then either $\beta$ is a desirable non-simple 2-form or $\beta \wedge \beta = 0$ and a non-simple form is $\alpha + \beta$.  Q.E.D.

*Proof of the Lemma* **1.8**  Let us chose in accordance with lemmas **1.6**, **1.7** some nonzero elements $\alpha, \mu$ of ${}^{\pm}\Lambda^2$ such that $\alpha \wedge \alpha = 0$, $\mu \wedge \mu \neq 0$. They are linearly independent. Let us assume at first that $\alpha \wedge \mu \neq 0$. Let $\beta_k = \mu + k\alpha$ for some number $k$. One has $\alpha \wedge \beta_k = \alpha \wedge \mu \neq 0$ and there exist $k$ such that $\beta_k \wedge \beta_k = \mu \wedge \mu + 2k\alpha \wedge \mu = 0$. Then $\beta = \beta_k$

---

* On leave of absence from: The National Research Institute For Physical Technical and Radio-Technical Measurements (VNIIFTRI), Mendeleevo, Moscow Region, 141570, Russia

will be simple. Let us chose now arbitrary $\delta \in {}^{\pm}\Lambda^2$ linearly independent in total with $\alpha, \beta$. For arbitrary $m, n$ the triad $\alpha, \beta, \delta_{mn} = \delta + m\alpha + n\beta$ constitutes the basis of ${}^{\pm}\Lambda^2$. Since $\alpha \wedge \delta_{mn} = \alpha \wedge \delta + n\alpha \wedge \beta$, $\beta \wedge \delta_{mn} = \beta \wedge \delta + m\alpha \wedge \beta$, as long as $\alpha \wedge \beta \neq 0$ there exist such $m, n$ that for $\gamma = \delta_{mn}$ $\alpha \wedge \gamma = 0 = \beta \wedge \gamma$. We assume $m, n$ to be chosen in a such way and fixed. The wedge square $\gamma \wedge \gamma$ must be nonzero otherwise $\gamma$ would be wedge orthogonal to all elements of ${}^{\pm}\Lambda^2$ contrary to the corollary **1.3**. The triad $\alpha, \beta, \gamma$ satisfy all the desirable conditions and thus in the case $\alpha \wedge \mu \neq 0$ the lemma implication is proved.

Now assume $\alpha \wedge \mu = 0$. Let $\nu$ be an arbitrary ${}^{\pm}\Lambda^2$ element independent in total with $\alpha, \mu$. Let us define $\nu_p = \nu + p\mu$ for some number $p$. $\nu_p \neq 0$ for every $p$. We have $\nu_p \wedge \mu = \nu \wedge \mu + p\mu \wedge \mu$ and since $\mu \wedge \mu \neq 0$ $\nu_p \wedge \mu = 0$ for some $p$. Let $\lambda = \nu_p$ for such a $p$. Since $\alpha \wedge \alpha = 0 = \alpha \wedge \mu$ $\alpha \wedge \lambda \neq 0$ otherwise a contradiction with corollary **1.3** would arise. Let $\lambda_q = \lambda + q\alpha$ for some number $q$. $\lambda_q \neq 0$ for every $q$. Further, $\alpha \wedge \lambda_q = \alpha \wedge \lambda \neq 0$, $\mu \wedge \lambda_q = \mu \wedge \lambda + q\mu \wedge \alpha = 0$ identically but $\lambda_q \wedge \lambda_q = \lambda \wedge \lambda + 2q\alpha \wedge \lambda$ depends on $q$ and can be zeroed by appropriate choice of it. Then the triad $\alpha, \beta = \lambda_q, \gamma = \mu$ will obey all the relations (1.1). Q.E.D.

*Proof of the Theorem* **1.11** Let $\alpha, \beta, \gamma$ possess all the properties (1.1), (1.2). Because $\alpha, \beta$ are simple and $\alpha \wedge \beta \neq 0$ there exist 1-forms $\tilde{\theta}^j$, $j = 1, 2, 3, 4$ such that $\alpha = \tilde{\theta}^1 \wedge \tilde{\theta}^2, \beta = \tilde{\theta}^3 \wedge \tilde{\theta}^4$. They are independent in total and form the basis of $\Lambda$. Let us carry out an expansion of $\gamma$ with respect to the basis $\tilde{\theta}^j \wedge \tilde{\theta}^k$ of $\Lambda^2$. The equations $\alpha \wedge \gamma = 0 = \beta \wedge \gamma$ restrict this expansion to the form

$$\gamma = (a\tilde{\theta}^1 + c\tilde{\theta}^2) \wedge \tilde{\theta}^3 + (b\tilde{\theta}^1 + d\tilde{\theta}^2) \wedge \tilde{\theta}^4.$$

The condition $\gamma \wedge \gamma \neq 0$ then yields $\Delta \equiv ad - bc \neq 0$ and it follows from (1.2) that $\Delta = 1/4$. This means that the transformation $\theta^1 = 2(a\tilde{\theta}^1 + c\tilde{\theta}^2)$, $\theta^2 = 2(b\tilde{\theta}^1 + d\tilde{\theta}^2)$ is reversible and hence the tetrad $\theta^1, \theta^2, \tilde{\theta}^3, \tilde{\theta}^4$ is also the basis of $\Lambda$. Then equation $\Delta = 1/4$ yields $\alpha = \theta^1 \wedge \theta^2$ and (1.4) implies that $2\gamma = \theta^1 \wedge \tilde{\theta}^3 + \theta^2 \wedge \tilde{\theta}^4$. In accordance with definition of $\tilde{\theta}^j$ $\beta = \tilde{\theta}^3 \wedge \tilde{\theta}^4$.

Let us introduce the spinor indexing of the tetrad and the basis of undotted lobe as follows
$$\theta_{0\dot{0}} = \theta^1, \quad \theta_{0\dot{1}} = \theta^2, \quad \theta_{1\dot{0}} = -\tilde{\theta}^4, \quad \theta_{1\dot{1}} = \tilde{\theta}^3;$$
$$S_0 \equiv S_{00} = \alpha, \quad S_1 \equiv S_{01} = S_{10} = \gamma, \quad S_2 \equiv S_{11} = \beta.$$

Then the equations (1.3$a$) are nothing else but the compact record of the relations between the bases of $\Lambda$ and ${}^{+}\Lambda^2$ listed above. Three other 2-forms expressed in terms of $\theta_{A\dot{B}}$ in accordance with (1.3$b$), i.e.
$$S_{\dot{0}} \equiv S_{\dot{0}\dot{0}} = \theta_{0\dot{0}} \wedge \theta_{1\dot{0}}, \quad S_{\dot{1}} \equiv S_{\dot{0}\dot{1}} = S_{\dot{1}\dot{0}} = \tfrac{1}{2}\theta_{0\dot{1}} \wedge \theta_{1\dot{0}} + \tfrac{1}{2}\theta_{0\dot{0}} \wedge \theta_{1\dot{1}}, \quad S_{\dot{2}} \equiv S_{\dot{1}\dot{1}} = \theta_{0\dot{1}} \wedge \theta_{1\dot{1}},$$
are, at first, linearly independent themselves and in total with $S_{AB}$ and, at second, wedge-orthogonal to the basis $S_{AB}$ of the undotted lobe. Thus they constitute the basis of the dotted lobe. The theorem is proven. Q.E.D.

*Proof of the Lemma* **1.15** Let $S_{AB}$ and $\tilde{S}_{AB}$ be two $S$-bases of the undotted lobe. The following expansion holds:
$$\tilde{S}_1 = aS_1 + bS_0 + cS_2.$$
Let us assume that $b \neq 0 \neq c$. We replace $S_{AB}$ by $g_0(\rho)S_{AB}$ with some number $\rho$. Then

$$\tilde{S}_1 = (b - a\rho + c\rho^2)g_0 S_0 + (a - 2\rho c)g_0 S_1 + cg_0 S_2$$

and one may choose such $\rho$ that the $g_0 S_0$−term vanishes. Thus there exists a rotation (possibly trivial) such that the expansion of $\tilde{S}_1$ with respect to rotated basis will not contain 0− or 2−suffixed terms. Furthermore, the application, if necessary, of the discrete rotation (1.8$d$) ensures exactly the 0−suffixed term will be absent. But since $\tilde{S}_1 \wedge \tilde{S}_1 \neq 0$ the coefficient of 1−suffixed term will be necessarily nonzero. Then the rotation $g_2$ can be applied to liquidate the 2−suffixed term as well. We see therefore that there exists such a rotation $\tilde{g} \in G'$ that $\tilde{S}_1 = \tilde{h}\tilde{g}S_1$ where $\tilde{h}$ is some nonzero number. Then 1−suffixed terms will be absent in the expansions of $\tilde{S}_0$ and $\tilde{S}_2$ with respect to basis $\tilde{g}S_{AB}$. Applying, if necessary, discrete rotation, one can ensure exactly 0−suffixed term to be nonzero in the $\tilde{S}_0$ expansion. Then the equation $\tilde{S}_0 \wedge \tilde{S}_1 = 0$ yields $\tilde{S}_0 = h_0 g S_0$ for some nonzero $h_0$ ($g$ equals $\tilde{g}$ or $g_\uparrow \tilde{g}$). At the same time $\tilde{S}_1 = h_1 \tilde{g} S_1$. Using these relations one easily obtains $\tilde{S}_2 = h_2 \tilde{g} S_2$ for some nonzero $h_2$. The rotation $g_1$ enables one to make equal the (transformed) coefficients $h_0$ and $h_2$ to $h = (h_0 h_2)^{1/2}$ remaining $h_1$ unaffected. Then equation $\tilde{S}_0 \wedge \tilde{S}_2 + 2\tilde{S}_1 \wedge \tilde{S}_1 = 0$ yields $h_1^2 = h^2$. If $h_1 = h$ then $\tilde{S}_{AB}$ and $g_1 g S_{AB}$ coincides up to overall factor $h$. If $h_1 = -h$ then we beforehand apply the rotation $g_1(-1)$ together with *conformal reflection* $S_{AB} \to -S_{AB}$ and come to the previous case.

The 'dotted' case need not separate consideration of course. The lemma is therefore proven.                                           Q.E.D.

*Proof of the Lemma* **1.16**   Lemma **1.15** ensures that $S_{AB}$ and $\tilde{S}_{AB}$ are connected by the composition of the conformal transformation and rotations (1.8) Since they are fitted with the same dotted $S$-basis and due to eq. (1.7) the conformal transformation may be only a conformal reflection or a trivial one. It is evident that inversing the sign of one of two undotted $S$-bases and, accompanying this with the multiplication of the corresponding tetrad to imaginary unit, one may assume $S$-basis to be connected by a rotation: $\tilde{S}_{AB} = gS_{AB}$. It is straightforward to prove that the transformations (1.10) do not affect the dotted $S$-forms but generate exactly the rotations (1.8) of undotted ones, i.e.

$$\tilde{S}_{AB} = gS_{AB} = \tfrac{1}{2}\epsilon^{\dot{K}\dot{L}}(g\theta)_{A\dot{K}} \wedge (g\theta)_{B\dot{L}}$$

where the transformation $g$ of tetrad is constructed from elementary rotations (1.10) in the same way (and with the same parameter values) as the $S$-forms rotation, denoted by the same symbol, is constructed from (1.8). (There is certainly a minor abuse of the use of $g$ here.) The relations

$$\tilde{\theta}_{A\dot{B}} \wedge \tilde{\theta}_{C\dot{D}} = \epsilon_{AC} S_{\dot{B}\dot{D}} + \epsilon_{\dot{B}\dot{D}} \tilde{S}_{AB}$$
$$g\theta_{A\dot{B}} \wedge g\theta_{C\dot{D}} = \epsilon_{AC} S_{\dot{B}\dot{D}} + \epsilon_{\dot{B}\dot{D}} S_{AB}$$

then imply $\tilde{\theta}_{A\dot{B}} \wedge \tilde{\theta}_{C\dot{D}} = g\theta_{A\dot{B}} \wedge g\theta_{C\dot{D}}$ and it is a simple matter to show that $\theta_{A\dot{B}} = \varepsilon g\theta_{A\dot{B}}$ where $\varepsilon = 1$ or $\varepsilon = -1$. We have seen above that there is additional possibility of the multiplication of tetrad to the imaginary unit. At whole, one has all the four factors $1^{1/4}$. The lemma is proven.                                           Q.E.D.

*Proof of the Corollary* **1.18**   Let the undotted ($S_{AB}$) and dotted ($S_{\dot{A}\dot{B}}$) $S$-bases of transection are given. Due to $S$-basis definition there exists a tetrad $\theta_{A\dot{B}}$ such that $S_{\dot{A}\dot{B}}$ is represented in terms of it in accordance with (1.3b) Then the tetrad $\tilde{S}_{AB}$ that is expressed in terms of $\theta_{A\dot{B}}$ precisely by (1.3a) is an $S$-basis of the undotted lobe. $\tilde{S}_{AB}$ obeys the conditions of the lemma **1.15**. Let $\tilde{S}_{AB} = kgS_{AB}$ for some nonzero $k$. Then the tetrad $g\theta_{A\dot{B}}$ will conformally fit $S_{AB}$ and $S_{\dot{A}\dot{B}}$. Q.E.D.

*Proof of the lemma* **1.19** In accordance with the corollary **1.18** $S$-bases $S_{AB}$ and $\tilde{S}_{\dot{A}\dot{B}}$ are fitted by some tetrad $\hat{\theta}_{A\dot{B}}$. Then $S_{AB}$ and $\tilde{S}_{AB}$ are conformally fitted with the same dotted $S$-basis by the tetrads $\hat{\theta}_{A\dot{B}}$ and $\tilde{\theta}_{A\dot{B}}$ respectively. The lemma 1.16 ensures $\hat{\theta}_{A\dot{B}}$ and $\tilde{\theta}_{A\dot{B}}$ to be connected by conformal transformation and undotted rotation. In a similar way $S_{\dot{A}\dot{B}}$ and $\tilde{S}_{\dot{A}\dot{B}}$ are conformally fitted with the same undotted $S$-basis by the tetrads $\theta_{A\dot{B}}$ and $\hat{\theta}_{A\dot{B}}$ respectively. Then they are connected by some conformal transformation and undotted rotation. As a result $\theta_{A\dot{B}}$ and $\tilde{\theta}_{A\dot{B}}$ are connected by a conformal transformation and rotation of the both types. Q.E.D.

*Proof of the theorem* **1.21** Let us show at first that the transection specifies the metric up to conformal factor. For, the transection equipped every lobe with a family of $S$-bases in accordance with theorem **1.11**. Due to the corollary **1.18** any pair of $S$-bases of different lobes is fitted by some tetrad $\theta_{A\dot{B}}$. Having any such tetrad, let us introduce the following symmetric (and obviously non-degenerated) second order tensor **g** by the definition

$$\mathbf{g} = \theta^A{}_{\dot{B}} \otimes \theta_A{}^{\dot{B}} \qquad (A.1)$$

It is easy to see that it does not depend on the arbitrariness involved in its definition (apart of the conformal rescaling). Indeed, due to the lemma **1.19** and the corollary **1.20** the tetrad $\theta_{A\dot{B}}$ is specified by transection up to a conformal factor and possibility of undotted and dotted rotations alone. It is straightforward to check that rotations do not affect the expression $(A.11)$ which therefore is determined up to a conformal transformation. The first implication of the theorem is proven.

To prove the inverse one we have to show that the metric specifies a unique transection.

Let us notice at first that any non-degenerate symmetric tensor **g** can me reduced (over $\mathbb{C}$) to a diagonal form

$$\mathbf{g} = \Theta^a \otimes \Theta^a, \qquad a = 1, 2, 3, 4 \qquad (A.2)$$

where 1-forms $\Theta^a$ constitute a basis of $\Lambda$. The corresponding spinor indexing may be introduced by the transition to tetrad $\theta_{A\dot{B}}$ in the following way:

$$\theta_{0\dot{0}} = \Theta^3 + i\Theta^4, \qquad \theta_{1\dot{1}} = -\Theta^3 + i\Theta^4,$$
$$\theta_{0\dot{1}} = \Theta^1 + i\Theta^2, \qquad \theta_{1\dot{0}} = -\Theta^1 + i\Theta^2.$$

Then the eq. (1.11) turns out to be equivalent to (1.12) and the formulae (1.3) yield the $S$-basis of undotted and dotted lobes of some transection. We must to prove that this transection is unique one obtainable in such a way.

In is straightforward to verify that the lobes of transection constructed above from the tetrad $\Theta^a$ which is orthonormal with respect to **g** are at the same time linear spaces spanned by the following two collections of 2-forms

$$\Sigma^{ab} = \Theta^a \wedge \Theta^b + \tfrac{1}{2}\varepsilon^{abcd}\Theta^c \wedge \Theta^d \qquad (A.3a)$$
$$\dot{\Sigma}^{ab} = \Theta^a \wedge \Theta^b - \tfrac{1}{2}\varepsilon^{abcd}\Theta^c \wedge \Theta^d \qquad (A.3b)$$

There are three distinct elements in every collection $(A.3)$ in fact.

Let us notice now that the expansion $(A.2)$ of conformal metric specifies the tetrad up to conformal factor and complex orthogonal $O(4, \mathbb{C})$ transformations. With respect to conformal transformations linear spaces spanned by the 2-forms collections $(A.3)$ are evidently invariant. Let us consider the effect of orthogonal transformations.

Latter can be proper or non-proper one. A non-proper rotation can be in turn expanded to superposition of a proper one and reflection of some direction which always can be chosen as the element of the basis. But the reflection of, say, $\Theta^2$ corresponds precisely to the interchange of lobes and does not affect the transection itself. Thus the problem is reduced to the consideration of proper rotations only.

Let the matrix $L^{ab} \in SO(4, \mathbb{C})$. The second rotated tetrad $\tilde{\Theta}^a = L^{ab}\Theta^b$ yields a new 'undotted' family of 2-forms in the same way as initial one:
$$\tilde{\Sigma}^{ab} = L^{ak}L^{bl}\Theta^k \wedge \Theta^l + \tfrac{1}{2}\varepsilon^{abcd}L^{ck}L^{dl}\Theta^k \wedge \Theta^l \qquad (A.4)$$

Our goal is to prove that $\tilde{\Sigma}^{ab} \wedge \dot{\Sigma}^{kl} = 0$ *identically*. Due to lemma **1.2** this would mean that the collections of 2-forms $(A.4)$ and $(A.3a)$ span the same subspace. Since $L^{ab}$ is arbitrary orthogonal matrix, the invariance of this subspace (and transection as whole) will follow.

One has $\Theta^a \wedge \Theta^b \wedge \Theta^c \wedge \Theta^d = \varepsilon^{abcd}\Theta^1 \wedge \Theta^2 \wedge \Theta^3 \wedge \Theta^4$ and hence it is necessary to prove that
$$\mathbf{L} \equiv \varepsilon^{klmn}L^{ak}L^{bl} - \tfrac{1}{2}\varepsilon^{mncd}\varepsilon^{klcd}L^{ak}L^{ll} + \tfrac{1}{2}\varepsilon^{abcd}\varepsilon^{klmn}L^{ck}L^{dl}$$
$$- \tfrac{1}{4}\varepsilon^{mnpq}\varepsilon^{abcd}\varepsilon^{pqkl}L^{ck}L^{dl} = 0$$

The identity $\varepsilon^{abcd}\varepsilon^{klmn} = 2\delta^{ak}\delta^{bl} - 2\delta^{al}\delta^{bk}$ yields
$$\mathbf{L} = \varepsilon^{klmn}L^{ak}L^{bl} - L^{am}L^{bn} + L^{an}L^{bm} + \tfrac{1}{2}\varepsilon^{klmn}\varepsilon^{abcd}L^{ck}L^{dl} - \varepsilon^{abcd}L^{cm}L^{dn}.$$
The orthogonality of $L^{ab}$ is equivalent to equality $\delta^{ab} = L^{ca}L^{cb}$. Hence
$$\varepsilon^{abcd}L^{mc}L^{nd} = L^{pa}L^{pr}L^{qb}L^{qs}\varepsilon^{rscd}L^{mc}L^{nd}$$

But $\varepsilon^{rscd}L^{pr}L^{qs}L^{mc}L^{nd} = \varepsilon^{pqmn}\det L = \varepsilon^{pqmn}$ and hence $\varepsilon^{abcd}L^{mc}L^{nd} = \varepsilon^{mnpq}L^{pa}L^{qb}$. Similarly $\varepsilon^{abcd}L^{cm}L^{cn} = \varepsilon^{mnpq}L^{ap}L^{bq}$. Using these identities one can easily deduce from the above $\mathbf{L}$ representation that $\mathbf{L} \equiv 0$.

We have shown that $\Sigma^{ab}$ and $\tilde{\Sigma}^{ab}$ span the (3-dimensional) subspaces that are wedge orthogonal to the same subspace spanned by $\dot{\Sigma}^{ab}$. Then lemma **1.2** implies that subspaces spanned by $\Sigma^{ab}, \tilde{\Sigma}^{ab}$ coincide.

In a similar way one may prove that rotation of the tetrad $\Theta^a$ does not affect the subspace spanned by the $(A.3b)$. The theorem is proven. Q.E.D.

*Proof of the corollary* **1.22** It follows immediately from the corresponding invariance of 2-forms $(A.3)$ and the above theorem proof. Q.E.D.

*Proof of the proposition* **2.4** In accordance with corollary **1.18** any $S$-basis of (say) undotted lobe is conformally fitted with any $S$-basis of dotted lobe. If transection is normalized, the conformal factor is reduced to one of to values 1 or -1. In the former case $S$-bases are fitted. On the other hand it is easy to see that if $S$-bases $S_{AB}$ and $S_{\dot{A}\dot{B}}$ are fitted (i.e. formulae $(1.3)$ holds for some tetrad) then $S_{AB}$ and $-S_{\dot{A}\dot{B}}$ cannot be fitted. Q.E.D.

*Proof of the lemma* **3.3** (proposition **3.2** is proven in the next item). Let us choose arbitrary $S$-basis $\tilde{S}_{AB}$ of the lobe and expand $\alpha$ with respect to it: $\alpha = a\tilde{S}_0 + b\tilde{S}_1 + c\tilde{S}_2$. Let us assume $C \neq 0$. If $a \neq 0$ one may apply $g_2(\rho)$ rotation choosing $\rho$ to be a root of the equation $a\rho^2 - b\rho + c = 0$. Then $\alpha = ag_2\tilde{S}_0 + \tilde{b}g_2\tilde{S}_1$ where $\tilde{b} = b - 2a\rho$. On the other hand if $a = 0$ then $\alpha = cg_\uparrow\tilde{S}_0 - bg_\uparrow\tilde{S}_1$. Thus in any case there exists a rotation $\tilde{g}$ such that $\alpha = \tilde{a}\tilde{g}\tilde{S}_0 + b'\tilde{g}\tilde{S}_1$. Then if $\alpha \wedge \alpha = 0$ it follows immediately that $b' = 0$ but $\tilde{a} \neq 0$ and rotation $g_1(\tilde{a}^{1/2})$ yields $\alpha = g_1\tilde{g}\tilde{S}_0$. In another case $\alpha \wedge \alpha \neq 0$ one has $b' \neq 0$ and $\alpha = b'g_0(\tilde{a}/b')\tilde{g}\tilde{S}_0$, so $k = b'$ and $S = g_0\tilde{g}S$. The lemma is proven. Q.E.D

*Proof of the proposition* **3.2** A necessity follows from the lemma **3.3**. Indeed, one may always find an $S$-basis such that $\alpha = S_0$. Then $\beta = S_2$ is a simple lobe element obeying the relation $\alpha \wedge \beta \neq 0$.

Now we shall prove a sufficiency. Let $\alpha, \beta$ be any $^?\Lambda^2$ simple elements with non-vanishing wedge product. There is a tetrad $\Theta^j \in \Lambda$, $j = 1, 2, 3, 4$ such that $\alpha = \Theta^1 \wedge \Theta^2, \beta = \Theta^3 \wedge \Theta^4$. Let $\tilde{\gamma}$ be any other $^?\Lambda^2$ element linearly independent in total with $\alpha, \beta$. The triad $\{\alpha, \beta, \tilde{\gamma}\}$ constitutes a basis of $^?\Lambda^2$. If $\tilde{\gamma} \wedge \tilde{\gamma} = 0$ there exist (due to proposition conditions) a simple 2-form $\delta$ such that $\tilde{\gamma} \wedge \delta \neq 0$. Since $\delta$ is simple the expansion $\delta = a\alpha + b\beta + c\tilde{\gamma}$ implies $c \neq 0$. Let $\gamma = \delta + h\tilde{\gamma}$. Then with $h \neq -c$ the triad $\{\alpha, \beta, \gamma\}$ constitutes basis of $^?\Lambda^2$ as well and for $h \neq 0$ one has $\gamma \wedge \gamma \neq 0$. We have proven therefore that there always exists element $\gamma \in {}^?\Lambda^2$ such that $\gamma \wedge \gamma \neq 0$ and the triad $\{\alpha, \beta, \gamma\}$ is a basis of $^?\Lambda^2$.

Let us define now $\gamma' = \gamma + k\alpha + l\beta$ for some numbers $k, l$. Since $\alpha \wedge \beta \neq 0$ it is clear that one may enforce $\alpha \wedge \gamma'$ and $\beta \wedge \gamma'$ to vanish by means of their proper choice. Linear independence of $\alpha, \beta, \gamma$ guarantees that $\gamma'$ will be nonzero and, moreover, the triad $\{\alpha, \beta, \gamma'\}$ constitutes a basis of $^?\Lambda^2$. Due to proposition conditions if $\gamma' \wedge \gamma' = 0$ there exist a simple $\delta'$ obeying $\gamma' \wedge \delta' \neq 0$. But as well as any $^?\Lambda^2$ element $\delta'$ has to admit an expansion of the form $\delta' = a'\alpha + b'\beta + c'\gamma'$. This however is impossible because it would contradict the above assumptions concerning the wedge product values. Thus $\gamma' \wedge \gamma' \neq 0$.

It follows from representation of $\alpha$ and $\beta$ in terms of tetrad $\Theta^j$ that

$$\gamma' = \Theta^1 \wedge (k'\Theta^3 + l'\Theta^4) + \Theta^2 \wedge (m'\Theta^3 + n'\Theta^4)$$

and $\Delta \equiv k'n' - l'm' \neq 0$. Without loss of generality one may assume $\Delta = 1$. Then introducing $\hat{\Theta}^3 = k'\Theta^3 + l'\Theta^4, \hat{\Theta}^4 = m'\Theta^3 + n'\Theta^4$ one obtains $\alpha = \Theta^1 \wedge \Theta^2, \beta = \hat{\Theta}^3 \wedge \hat{\Theta}^4, \gamma' = \Theta^1 \wedge \hat{\Theta}^3 + \Theta^2 \wedge \hat{\Theta}^4$. Hence the 2-forms $S_0 = \alpha, S_1 = -\frac{1}{2}\gamma', S_2 = \beta$ constitutes the basis of the space spanned by them. Furthermore, the triad $S_{\dot{0}} = \Theta^1 \wedge \hat{\Theta}^4, S_{\dot{1}} = -\frac{1}{2}\Theta^1 \wedge \hat{\Theta}^3 - \frac{1}{2}\Theta^2 \wedge \hat{\Theta}^4, S_{\dot{0}} = \hat{\Theta}^3 \wedge \Theta^2$ is the $S$-basis of the complement $^-\Lambda^2$ of $^+\Lambda^2 = {}^?\Lambda^2$ in $\Lambda^2$. The triads $S_{AB}$ and $S_{\dot{A}\dot{B}}$ span 3-dimensional subspaces of $\Lambda^2$ that are easy to see to possess all the properties of complementary transection lobes. The proposition is proven. Q.E.D

*Proof of the theorem* **3.5** Direct implication. Let us assume that 2-dimensional subspace of a lobe contains only simple elements. Let $\alpha$ and $\beta$ be such linearly independent 2-forms. Lemma **3.3** implies an existence of $S$-basis such that $\alpha = S_0$. Let $\beta = aS_0 + bS_1 + cS_2$. Since $\alpha \wedge \alpha = 0 = \beta \wedge \beta$  $\alpha + \beta$ will be a simple form if and only if $\alpha \wedge \beta = 0$. But the latter requirement implies $c = 0$ and then simplicity of $\beta$ will yield $b = 0$. This is impossible because $\alpha$ and $\beta$ cannot be proportional. Thus $\alpha \wedge \beta \neq 0$ and $\alpha + \beta$ is a desirable non-simple 2-form.

Reversed implication. Let $\alpha$ be a simple nonzero element of a 3-dimensional subspace $^?\Lambda^2$ of $\Lambda^2$ and $\gamma$ any independent with $\alpha$ element of $^?\Lambda^2$. The subspace spanned by $\alpha, \gamma$ contain a non-simple element and one may assume that it coincides with $\gamma$. If $\alpha \wedge \gamma \neq 0$ there is a number $k$ such that $\beta = \alpha + k\gamma$ is the simple (and nonzero) $^?\Lambda^2$ element. At the same time $\alpha \wedge \beta = \alpha \wedge \gamma \neq 0$ and therefore we have found a simple element of $^?\Lambda^2$ which is not wedge-orthogonal to $\alpha$.

The second possibility is $\alpha \wedge \gamma = 0$. Let us choose an element $\tilde{\beta} \in {}^?\Lambda^2$ independent in total with $\alpha, \gamma$. Since $\gamma \wedge \gamma \neq 0$ a (nonzero) 2-form $\beta = \tilde{\beta} + l\gamma$ is simple for some $l$. But $\alpha \wedge \beta \neq 0$ since otherwise $\alpha$ and $\beta$ would span a subspace which consists of a simple elements only.

We see that in both cases the conditions of the theorem 3.2 are fulfilled and therefore $^?\Lambda^2$ is a lobe of some transection. Q.E.D

*Proof of the lemma* **3.8** Let us assume that it is not the case and let $\tilde{\Lambda}$ be more than 3-dimensional subspace of $\Lambda^2$ without simplest subspaces. Let us also choose any 3-dimensional subspace $^+\Lambda^2$ of $\tilde{\Lambda}$. It obeys the conditions of the theorem **3.5'** and is therefore a lobe of some transection $^+\Lambda^2 \oplus {}^-\Lambda^2 = \Lambda^2$. Since the dimension of $\tilde{\Lambda}$ exceeds 3 there exists $\alpha \in \tilde{\Lambda}$ such that $\alpha \notin {}^+\Lambda^2$. One may assume $\alpha = \alpha^+ + \alpha^-$, $\alpha^\pm \in {}^\pm\Lambda^2$ and $\alpha^- \neq 0$. Since $^+\Lambda^2 \subset \tilde{\Lambda}$, $\alpha^- = \alpha - \alpha^+ \in \tilde{\Lambda}$. One has either $\alpha^- \wedge \alpha^- = 0$ or $\alpha^- \wedge \alpha^- \neq 0$. In the former case we take $\beta = \alpha^-$ and $\gamma$ to be any simple nonzero element of $^+\Lambda^2$. In the latter case there exists $\tilde{\alpha} \in {}^+\Lambda^2$ such that $\beta = \tilde{\alpha} + \alpha^-$ is simple (it is evident that $\tilde{\alpha}$ may be any properly scaled *non-simple* $^+\Lambda^2$ element). It is easy to prove using lemma **3.3** that there is a simple nonzero $\gamma \in {}^+\Lambda^2$ such that $\gamma \wedge \alpha = 0$.

In the both cases $\beta$ add $\gamma$ are linearly independent and span a simplest subspace of $\tilde{\Lambda}$, a contradiction. The lemma is proven. Q.E.D

*Proof of the theorem* **3.11** We must establish a denseness only. We shall show that a neighborhood of any 3-dimensional subspace of $\Lambda^2$ which is *not* a lobe of some transection contains a subspace which *is* a lobe. At first, we shall classify such a peculiar 3-spaces $\Lambda'$. In accordance with theorem **3.5'** such a space contains a simplest 2-space or is a simplest space itself. If a 3-dimensional subspace of $\Lambda^2$ is simplest it is spanned either by

(i) $\alpha = \Theta^1 \wedge \Theta^2, \beta = \Theta^1 \wedge \Theta^3, \gamma = \Theta^1 \wedge \Theta^4$ or by

(ii) $\alpha = \Theta^1 \wedge \Theta^2, \beta = \Theta^1 \wedge \Theta^3, \gamma = \Theta^2 \wedge \Theta^3$

where independent 1-forms $\Theta^j$ constitute the basis of $\Lambda$. If a simplest subspace is 2-dimensional it is spanned by $\alpha = \Theta^1 \wedge \Theta^2$ and $\beta = \Theta^1 \wedge \Theta^3$. Then let $\gamma$ be any element of $\Lambda'$ independent in total with $\alpha, \beta$. If

(iii) $\gamma$ is wedge orthogonal to both $\alpha$ and $\beta$

then it has to be non-simple. Then let us consider expansion of $\gamma$ with respect to basis $\Theta^i \wedge \Theta^j$. Subtracting if necessary terms proportional to $\alpha, \beta$ one may assume without loss of generality that the expansion does not contain $\Theta^1 \wedge \Theta^2$- and $\Theta^1 \wedge \Theta^3$-terms. The wedge orthogonality with $\alpha, \beta$ implies the absence of $\Theta^3 \wedge \Theta^4$- and $\Theta^2 \wedge \Theta^4$-terms as well. Then since $\gamma \wedge \gamma \neq 0$ both other terms proportional to $\Theta^2 \wedge \Theta^3$ and $\Theta^1 \wedge \Theta^4$ are present with nonzero coefficients. After rescaling of $\gamma$ and $\Theta^4$ one will obtain, without loss of generality, the following representation for the third (and last) element of $\Lambda'$ basis $\gamma = \Theta^2 \wedge \Theta^3 + \Theta^1 \wedge \Theta^4$.

The next possibility is the non-vanishing of either product $\gamma \wedge \beta$ or $\gamma \wedge \alpha$ or both together. In latter case however one may replace the pair $\alpha, \beta$ by some their linear combination (that implies linear transformation of $\Theta^2, \Theta^3$) in such a way that one of the above products will vanish. Thus without loss of generality one may assume $\gamma \wedge \beta \neq 0 = \gamma \wedge \alpha$. Furthermore, subtracting if necessary from $\gamma$ the element $\beta$ with appropriate coefficient one may reduce $\gamma$ to be simple. Hence we may additionally assume $\gamma \wedge \gamma = 0$.

Since $\alpha$ is simple and wedge orthogonal to $\beta, \gamma$ it can be added with arbitrary coefficient to $\gamma$. As a result we may assume that expansion of $\gamma$ with respect to basis $\Theta^i \wedge \Theta^j$ does not contain $\Theta^1 \wedge \Theta^2$-term. The equation $\alpha \wedge \gamma = 0$ then implies its following form:
$$\gamma = \Theta^1 \wedge [(\cdot)\Theta^3 + (\cdot)\Theta^4] + \Theta^2 \wedge [(\cdot)\Theta^3 + (\cdot)\Theta^4].$$
Since $\beta \wedge \gamma = \Theta^1 \wedge \Theta^3 \wedge \gamma \neq 0$ the very last above coefficient $(\cdot)$ is nonzero and after possible absorption of $\Theta^3$-term in the same brackets and rescalings one obtains $\gamma = \Theta^1 \wedge [(\cdot)\Theta^3 + (\cdot)\Theta^4] + \Theta^2 \wedge \Theta^4$. The last restriction $\gamma \wedge \gamma = 0$ yields $\gamma = (\cdot)\Theta^1 \wedge \Theta^4 + \Theta^2 \wedge \Theta^4$.

As a result we obtain the last two different 'canonical' representations of the third basis element:

(iv) $\gamma = \Theta^2 \wedge \Theta^4$ and
(v) $\gamma = (\Theta^1 + \Theta^2) \wedge \Theta^4$.

Now we have a complete classification (cases (i)-(v)) of all possible 3-subspaces of $\Lambda^2$ possessing a simplest subspace and are able to show that in every case such a subspace has in its any vicinity (in the Grassmann manifold) another 3-dimensional subspace which does not admit simplest subspaces. The nearness of elements of Grassmann manifold (3-subspaces of $\Lambda^2$) may be understood here as possibility to point out such their bases that are element-by-element close in the topology of $\Lambda^2$.

Let us consider existing possibilities one-by-one.

(i) $\Lambda'$ is spanned by 2-forms $\alpha = \Theta^1 \wedge \Theta^2, \beta = \Theta^1 \wedge \Theta^3, \gamma = \Theta^1 \wedge \Theta^4$. Let
$$\alpha_\varepsilon = \Theta^1 \wedge \Theta^2 + \varepsilon \Theta^3 \wedge \Theta^4, \quad \beta_\varepsilon = \Theta^1 \wedge \Theta^3 - \varepsilon \Theta^2 \wedge \Theta^4, \quad \gamma_\varepsilon = \Theta^1 \wedge \Theta^4 + \varepsilon \Theta^2 \wedge \Theta^3$$
where $\varepsilon$ is arbitrarily small number. Then the only following pair wedge products revive
$$\alpha_\varepsilon \wedge \alpha_\varepsilon = \beta_\varepsilon \wedge \beta_\varepsilon = \gamma_\varepsilon \wedge \gamma_\varepsilon = 2\varepsilon\omega \qquad \text{where} \quad \omega = \Theta^1 \wedge \Theta^2 \wedge \Theta^3 \wedge \Theta^4 \neq 0.$$
Hence if a 2-form $A = a\alpha_\varepsilon + b\beta_\varepsilon + c\gamma_\varepsilon$ is simple then $a^2 + b^2 + c^2 = 0$, i.e. $A = a\alpha_\varepsilon + b\beta_\varepsilon + i(a^2 + b^2)^{1/2}\gamma_\varepsilon$ for some $a, b$ and some choice of the branch of square root. Let now $A_1$ and $A_2$ be two such *distinct* simple 2-forms. Their wedge orthogonality is equivalent to the equation $a_1 a_2 + b_1 b_2 = (a_1^2 + b_1^2)^{1/2}(a_2^2 + b_2^2)^{1/2}$. But it implies $a_1 b_2 - b_1 a_2 = 0$ that allows to show that $A_1$ and $A_2$ are proportional, a contradiction. Thus we see that any two different (non-proportional) simple 2-forms constructed from $\alpha_\varepsilon, \beta_\varepsilon, \gamma_\varepsilon$ are not wedge orthogonal. The conditions of the theorem **3.2** are fulfilled and the space spanned by $\alpha_\varepsilon, \beta_\varepsilon, \gamma_\varepsilon$ is a lobe of some transection. On the other hand for sufficiently small $\varepsilon$ it will belong to any preliminary chosen vicinity of original $\Lambda'$.

Now let us consider other possibilities (ii-v).

(ii) Now $\alpha = \Theta^1 \wedge \Theta^2, \beta = \Theta^1 \wedge \Theta^3, \gamma = \Theta^2 \wedge \Theta^3$ and let us take
$$\alpha_\varepsilon = \Theta^1 \wedge \Theta^2 + \varepsilon \Theta^3 \wedge \Theta^4, \quad \beta_\varepsilon = \Theta^1 \wedge \Theta^3 - \varepsilon \Theta^2 \wedge \Theta^4, \quad \gamma_\varepsilon = \Theta^2 \wedge \Theta^3 + \varepsilon \Theta^1 \wedge \Theta^4.$$

The 2-forms $\alpha_\varepsilon, \beta_\varepsilon, \gamma_\varepsilon$ possess the same properties with respect to wedge multiplication as in the case (i) and therefore the further proof is the same as above.

(iii)  $\Lambda'$ is spanned by $\alpha = \Theta^1 \wedge \Theta^2, \beta = \Theta^1 \wedge \Theta^3, \gamma = \Theta^1 \wedge \Theta^4 + \Theta^2 \wedge \Theta^3$. We define
$$\alpha_\varepsilon = \Theta^1 \wedge \Theta^2 + \varepsilon\Theta^3 \wedge \Theta^4, \quad \beta_\varepsilon = \Theta^1 \wedge \Theta^3 - \varepsilon\Theta^2 \wedge \Theta^4, \quad \gamma_\varepsilon = \gamma.$$
Then
$$\alpha_\varepsilon \wedge \alpha_\varepsilon = \beta_\varepsilon \wedge \beta_\varepsilon = 2\varepsilon\omega, \quad \gamma_\varepsilon \wedge \gamma_\varepsilon = -\omega.$$

All the other wedge products vanish. This list is rather similar to one encountered in the case (i) and that proof suits *mutatis mutandis* here as well.

(iv)  $\Lambda'$ is spanned by $\alpha = \Theta^1 \wedge \Theta^2, \beta = \Theta^1 \wedge \Theta^3, \gamma = \Theta^2 \wedge \Theta^4$. Let
$$\alpha_\varepsilon = \beta, \quad \beta_\varepsilon = \gamma, \quad \gamma_\varepsilon = \alpha + \varepsilon\Theta^3 \wedge \Theta^4.$$
This triad obeys the eq. (1.1) and is in fact $S$-basis (with $\gamma_\varepsilon$ rescaled). Thus $\alpha_\varepsilon, \beta_\varepsilon, \gamma_\varepsilon$ span a lobe of transection.

(v)  $\Lambda'$ is spanned by $\alpha = \Theta^1 \wedge \Theta^2, \beta = \Theta^1 \wedge \Theta^3, \gamma = (\Theta^1 + \Theta^2) \wedge \Theta^4$. Let
$$\alpha_\varepsilon = \Theta^1 \wedge \Theta^2 + \varepsilon\Theta^3 \wedge \Theta^4, \quad \beta_\varepsilon = \Theta^1 \wedge \Theta^3 - \varepsilon\Theta^2 \wedge \Theta^4, \quad \gamma_\varepsilon = (\Theta^1 + \Theta^2) \wedge \Theta^3 + \varepsilon\Theta^2 \wedge \Theta^3.$$
It is easy to see that the only nonzero products will be
$$\alpha_\varepsilon \wedge \alpha_\varepsilon = \beta_\varepsilon \wedge \beta_\varepsilon = \gamma_\varepsilon \wedge \gamma_\varepsilon = 2\varepsilon\omega, \quad \gamma_\varepsilon \wedge \beta_\varepsilon = -\omega.$$
Thus $(\beta_\varepsilon + k\gamma_\varepsilon) \wedge (\beta_\varepsilon + k\gamma_\varepsilon) = 2\varepsilon(k^2 - k/\varepsilon + 1)\omega$ and $\tilde{\alpha}_\varepsilon = \beta_\varepsilon + (2\varepsilon)^{-1}(1 + (1 - 4\varepsilon^2)^{1/2})\gamma_\varepsilon$, $\tilde{\beta}_\varepsilon = \beta_\varepsilon + (2\varepsilon)^{-1}(1 - (1 - 4\varepsilon^2)^{1/2})\gamma_\varepsilon$ are the simple elements whose wedge product does not vanish. Together with $\tilde{\alpha}_\varepsilon = \alpha$ they can be rearranged to form a $S$-basis and hence $\alpha_\varepsilon, \beta_\varepsilon, \gamma_\varepsilon$ span a lobe of transection.

All the cases have been considered, the theorem is proven.  Q.E.D.

*Proof of the theorem* **5.3**  At first we shall prove that $\Omega(\alpha)$, $\alpha \in \Lambda^2$ is specified by the definition **5.2** uniquely. Next we shall prove the pointwise linearity of the curvature map and finally that it smoothly depends on the point of manifold.

Let one of two possible orders of the lobes of oriented normalized transection field be chosen (*i.e.* it is decided which a lobe will be called 'undotted' and which a 'dotted' one) and let $\alpha \in {}^+\Lambda^2$, $\alpha_{|p} \neq 0$. Let further $\{\alpha \wedge \alpha\}_{|p} = 0$. The $S$-basis that satisfies condition $\alpha_{|p} = S_{0|p}$ is not fixed uniquely but only up to rotations that are reduced in $p$ to $g_0$. The latter constitutes a subgroup of undotted rotation group that preserves $S_0$ over $p$. These transformations being arbitrarily continued onto $p$ neighborhood does not affect *in p* the $\Omega_0$ component of the undotted curvature due to theorem **5.1** and therefore $\alpha_{|p} = \Omega_{0|p}$ is uniquely specified as well. In the case of simple $\alpha_{|p}$ the definition **5.2** is therefore correct.

Now let $\alpha$ be non-simple in $p$. The restriction $\alpha_{|p} = kS_{1|p}$ also does not specify $S$-basis uniquely. The corresponding arbitrariness consists of (i) the discrete rotation $g_\uparrow$ which causes $k$ to reverse a sign and (ii) rotations $g_1$ that keep $k$ unaffected. The union of these transformations arbitrarily but smoothly continued to vicinity of $p$ constitutes in accordance with (1.9) subgroup of the rotation group which yields all the $S$-bases obeying restriction $kS_{1|p} = \alpha_{|p}$. Due to the theorem **5.1** the curvature 2-forms are transformed in the same way as $S$-forms and hence the expression $\Omega^+(\alpha) = k\Omega_1$ remains unaffected under the action of mentioned subgroup.

Hence the map $\Omega$ is correctly defined on the whole undotted lobe $^+\Lambda^2$. For the dotted lobe a proof is essentially the same.

Next we shall prove the pointwise linearity of the curvature map. At first let us show that $\Omega(k\alpha) = k\Omega(\alpha)$ for any function $k$. Let $\alpha$ be non-simple in some domain. Whenever $k = 0$ or $k \neq 0$ in $p$ the equation $\Omega(k\alpha) = k\Omega(\alpha)$ in that point is satisfied directly due definition of $\Omega$. Now let $\{\alpha \wedge \alpha\}_{|p} = 0$. The linearity property is trivially implied by $\Omega$ definition in the points where $k = 0$ or $\alpha = 0$. Let us drop such points and assume $k_{|p} \neq 0 \neq \alpha_{|p}$. If the $S$-basis $S_{AB}$ obeys $\alpha_{|p} = kS_{0|p}$ then the 2-form $k\alpha$ corresponds in particular (in terms of the same sort relation) to the $S$-basis $\tilde{S} = g_0(k^{1/2})S$. Due to the theorem **5.1** the corresponding curvature 2-forms $\Omega_{0|p}$ and $\tilde{\Omega}_{0|p}$ that are equal to $\Omega^+(\alpha)_{|p}$ and $\Omega^+(k\alpha)_{|p}$ respectively in virtue of definition **5.2** are connected by equation $\Omega_{0|p} = g_0(k^{1/2})\tilde{\Omega}_{0|p}$. A further speculation is rather obvious and the desirable property $\Omega(k\alpha) = k\Omega(\alpha)$ may be considered to be established.

Now we shall prove an additivity property $\Omega(\alpha + \beta) = \Omega(\alpha) + \Omega(\beta)$ for the case $\alpha, \beta \in {}^+\Lambda^2$ (or $\alpha, \beta \in {}^-\Lambda^2$). For other possibilities the additivity directly follows from $\Omega$ definition or is reduced to a mentioned one. The main part of proof is absorbed by the following

*Technical lemma* Let $\alpha, \beta$ be linearly independent elements of the same lobe (for definiteness let it be $^+\Lambda^2$) over some point. Then there exist $S$-basis of this lobe such that in that point

(i) if $\alpha \wedge \alpha = 0 = \beta \wedge \beta$ then $\alpha = S_0, \beta = kS_2$ and
$$\alpha + \beta = 2ik^{1/2}g_0\bigl((2ik^{1/2})^{-1}\bigr)g_2(-ik^{1/2})S_1;$$

(ii) if $\alpha \wedge \alpha = 0, \beta \wedge \beta \neq 0, \alpha \wedge \beta = 0$ then $\alpha = S_0, \beta = kS_1$ and
$$\alpha + \beta = kg_0(k^{-1})S_1;$$

(iii) if $\alpha \wedge \alpha = 0, \beta \wedge \beta \neq 0, \alpha \wedge \beta \neq 0$ then $\alpha = S_0, \beta = kS_1 + lS_2$ and
$$\alpha + \beta = g_2(k/2)S_0 \quad \text{if } l = k^2/4 \qquad \text{otherwise}$$
$$\alpha + \beta = 2ag_0\bigl((2a)^{-1}\bigr)g_2(k/2 - a)S_1 \quad \text{where } a = (k^2/4 - l)^{1/2};$$

(iv) if $\alpha \wedge \alpha \neq 0, \beta \wedge \beta \neq 0, \alpha \wedge \beta = 0$ then $\alpha = kS_1, \beta = l(S_0 + S_2)$ and
$$\alpha + \beta = lg_2(\pm 1)S_0 \quad \text{if } k = \pm l \qquad \text{otherwise}$$
$$\alpha + \beta = ag_0(l/a)g_2\bigl((k-a)/(2l)\bigr)S_1, \quad a = (k^2 - 4l^2)^{1/2};$$

(v) if $\alpha \wedge \alpha \neq 0, \beta \wedge \beta \neq 0, \alpha \wedge \beta \neq 0$ then either $\alpha = kS_1, \beta = l(S_0 + S_1)$ and
$$\alpha + \beta = lS_0 \quad \text{if } k + l = 0 \text{ and}$$
$$\alpha + \beta = (k+l)g_0\bigl(l/(k+l)\bigr)S_1 \text{ otherwise} \qquad \text{or}$$

(vi) $\alpha = kS_1, \beta = m(S_0 + S_2)$ and
$$\alpha + \beta = mg_2(\pm 1)S_1 \quad \text{if } k + l = \pm 2m \text{ else}$$
$$\alpha + \beta = ag_0(m/a)g_2\bigl((k + l - a)/(2m)\bigr)S_1, \quad a = \bigl((k+l)^2 - 4m^2\bigr)^{1/2}.$$

Here $k, l, m$ are some nonzero in the point functions.

The proof of the technical lemma occupies the next item of Appendix and there we proceed with the proof of the theorem **5.3**.

We shall show the pointwise additivity of $\Omega$ in the case (i) only because the other cases are handled in a quite similar way.

Due to the relations to $S$-forms implied by the item (i) of technical lemma one has $\Omega(\alpha) = \Omega_0$. The $\beta$-form can be represented as $\beta = k g_\uparrow S_0$ and transformation properties of curvature forms (theorem **5.1**) imply $\Omega(\beta) = k\Omega(g_\uparrow S_0) = k g_\uparrow \Omega(S_0) = k g_\uparrow \Omega_0 = k\Omega_2$ so we have $\Omega(\alpha) + \Omega(\beta) = \Omega_0 + k\Omega_2$. On the other hand $\Omega(\alpha + \beta) = 2ik^{1/2}\Omega\left(g_0\big((2ik^{1/2})^{-1}\big)g_2(-ik^{1/2})S_1\right) = 2ik^{1/2}g_0 g_2 \Omega(S_1) = \Omega_0 + k\Omega_2$ i.e. $\Omega(\alpha + \beta) = \Omega(\alpha) + \Omega(\beta)$.

We may consider the pointwise linearity of the map $\Omega$ to be established.

Finally we shall prove that the curvature map smoothly depends on a point. The problem can arise because there may be the points in a neighborhood of a given one with a different curvature map calculation rule implied by definition **5.2**. But it is easy to see that the smoothness almost directly follows from pointwise linearity. Indeed, the definition **5.2** implies that for every undotted $S$-basis the values of $\Omega(S_{AB})$ are the smooth local fields of 2-forms. But every $\alpha \in {}^+\Lambda^2$ can be expanded with respect to $S$-basis and the equation holds $\alpha = A^{AB} S_{AB}$ where $A^{AB}$ are some smooth functions. Then the pointwise linearity of $\Omega$ implies $\Omega(\alpha) = A^{AB}\Omega(S_{AB})$ which is a smooth local field.

It is evident that the 'dotted' argument of the curvature map may be analyzed in exactly the same way and in the 'mixed' case the consideration is almost trivial. The theorem is proven. Q.E.D.

*Proof of the technical lemma* If $\alpha, \beta$ are expressed via $S$-forms in according with the lemma assertions then it is a matter of straightforward computation to verify the validity of $\alpha + \beta$ representations with the help of equations (1.8). It remains to establish the representations of $\alpha, \beta$ via $S$-basis.

Case (i). In accordance with lemma **3.3** there exist $S$-basis $\tilde{S}_{AB}$ such that $\alpha = \tilde{S}_0$. One has the following general expansion for $\beta$: $\beta = a\tilde{S}_2 + b\tilde{S}_1 + c\tilde{S}_0$. If $a = 0$ then the condition $\beta \wedge \beta = 0$ implies additionally $b = 0$ that would mean proportionality of $\alpha, \beta$, a contradiction. Thus $a \neq 0$. Let us transform $S$-basis $\tilde{S}_{AB}$ to $S = g_0(\rho)\tilde{S}$. Then $\beta = (c - \rho b + \rho^2)S_0 + (b - 2\rho a)S_1 + aS_2$. By the proper choice of $\rho$ the $S_0$-term may be removed. Then the condition $\beta \wedge \beta = 0$ yields the vanishing of the $S_1$-term as well and we obtain $\beta = aS_2$ as it was required.

Case (ii). As above one may assume $\alpha = \tilde{S}_0$. Then $\alpha \wedge \beta = 0$ implies $\beta = b\tilde{S}_1 + c\tilde{S}_2$ but since $\beta \wedge \beta \neq 0$ one has $b \neq 0$. The rotation $g_0(c/b)$ of the $S$-basis yields then a desirable result.

Case (iii). As above we have $\alpha = \tilde{S}_0$, $\beta = a\tilde{S}_2 + b\tilde{S}_1 + c\tilde{S}_0$ and since $\alpha \wedge \beta \neq 0$ $a \neq 0$. As in the case (i) the rotation $g_0(\rho)$ enables one to remove $\tilde{S}_0$-term. Then $S_1$-term will have nonzero coefficient and required representation follows.

Case (iv) Due to lemma **2.3** there is $S$-basis $S_{AB}$ such that $\alpha = kS_1$. Since $\alpha \wedge \beta = 0$ one has $\beta = a\tilde{S}_0 + b\tilde{S}_2 = (ab)^{1/2}\big((a/b)^{1/2}\tilde{S}_0 + (b/a)^{1/2}\tilde{S}_2\big)$. Then for $S = g_1\big((a/b)^{1/4}\big)\tilde{S}$ a required representation follows.

Cases (v),(vi). As above we have $\alpha = \tilde{S}_0, \beta = a\tilde{S}_2+b\tilde{S}_1+c\tilde{S}_0$. Since $\alpha\wedge\beta \neq 0$ $b \neq 0$. Besides, either $a$ or $c$ is nonzero. If $ac \neq 0$ the rotation $g_1\big((ac)^{1/4}\big)$ yields the expansion asserted in the case (vi). If $a = 0, c \neq 0$ the rotations $g_1 g_\uparrow$ yield the representation (v) and if $a \neq 0, c = 0$ it is achieved by rotation $g_1$. The lemma is proven. Q.E.D.

*Proof of the lemma* **6.1** Let us choose in accordance with local version of the lemma **3.3** undotted $S$-basis $S_{AB}$ such that $\alpha = S_0$. Let further $S_{\dot{A}\dot{B}}$ be the dotted $S$-basis fitted with $S_{AB}$. One has $\beta = aS_{\dot{0}} + bS_{\dot{1}} + cS_{\dot{1}}$ for some functions $a, b, c$. Due to equation $\beta\wedge\beta = 0$ $a$ and $c$ are both nonzero provided $b \neq 0$. By means of the dotted rotation $\dot{g}_2$ the $S_{\dot{2}}$-term in the above $\beta$ expression can be removed meanwhile $S_{\dot{0}}$-coefficient $a$ remains unchanged. Then it follows that $S_{\dot{1}}$-term vanishes as well. Further $S_{\dot{0}}$-coefficient is reduced to the unit by means of rotation $\dot{g}_1$ and a required equation $\beta = S_{\dot{0}}$ follows.

Now let us assume that $b = 0$ in some point. Then precisely one of functions $a, c$ is nonzero in that point. Due to possibility of discrete rotation $\dot{g}_\uparrow$ we may assume $a \neq 0$. Then we come to the previous case by means, for example, a rotation $\dot{g}_0$ with any nonzero argument.

It is clear that in both cases all the object can be chosen to depend smoothly on the point. The lemma is proven. Q.E.D.

*Proof of the theorem* **6.2** Direct implication. If **g** is a real metric of Lorentz signature then locally
$$\varepsilon'\mathbf{g} = \Theta^0 \otimes \Theta^0 - \Theta^1 \otimes \Theta^1 - \Theta^2 \otimes \Theta^2 - \Theta^3 \otimes \Theta^3$$
where $\varepsilon' = 1$ or $\varepsilon' = -1$, $\Theta^j$ is some (real) basis of $\Lambda_{\mathbb{R}}$. Let us denote
$$\theta_{0\dot{0}} = \frac{\varepsilon}{\sqrt{2}}(\Theta^0 + \Theta^1), \theta_{1\dot{1}} = \frac{\varepsilon}{\sqrt{2}}(\Theta^0 - \Theta^1), \theta_{0\dot{1}} = \frac{\varepsilon}{\sqrt{2}}(\Theta^2 + i\Theta^3), \theta_{1\dot{0}} = \frac{\varepsilon}{\sqrt{2}}(\Theta^2 - i\Theta^3),$$
where $\varepsilon = 1$ if $\varepsilon' = 1$ and $\varepsilon = i$ if $\varepsilon' = -1$ ($\theta_{A\dot{B}}$ is already the basis of complexified space $\Lambda$ of course). Then (1.11) is fulfilled. The corresponding $S$-forms may be defined by (1.3) and it is easy to see that they satisfy $\overline{S_{AB}} \subset {}^-\Lambda^2$, $\overline{S_{\dot{A}\dot{B}}} \subset {}^+\Lambda^2$ and constitute the basis of $\Lambda^2$. Thus the complex conjugation interchanges the lobes of constructed transection. An overall conformal factor cannot affect these relations. The transection is conjugating symmetrical.

Reverse implication. Let $S_{AB}$ be any $S$-basis of the lobe ${}^+\Lambda^2$. Then $\overline{S_0} \in {}^-\Lambda^2$ and $\overline{S_0} \wedge \overline{S_0} = 0$. In accordance with lemma **6.1** there exist dotted $S$-basis $S_{\dot{A}\dot{B}}$ and the tetrad $\theta_{A\dot{B}}$ fitting $S_{AB}$ and $S_{\dot{A}\dot{B}}$ such that $\overline{S_0} = S_{\dot{0}}$, i.e.
$$\overline{\theta_{0\dot{0}}} \wedge \overline{\theta_{0\dot{1}}} = \theta_{0\dot{0}} \wedge \theta_{1\dot{0}}.$$
The latter equation implies
$$\overline{\theta_{0\dot{0}}} = a\theta_{0\dot{0}} + b\theta_{1\dot{0}}, \qquad (A.5)$$
$$\overline{\theta_{0\dot{1}}} = c\theta_{0\dot{0}} + d\theta_{1\dot{0}}, \qquad (A.6)$$
and, moreover, $ad - cd = 1$. Solving these linear equations, one obtains
$$\left.\begin{array}{l}\theta_{0\dot{0}} = d\overline{\theta_{0\dot{0}}} - b\overline{\theta_{0\dot{1}}} \\ \theta_{1\dot{0}} = a\overline{\theta_{0\dot{0}}} - c\overline{\theta_{0\dot{1}}}\end{array}\right\} \Rightarrow \begin{array}{ll}\overline{\theta_{0\dot{0}}} = a\theta_{0\dot{0}} + b\theta_{1\dot{0}}, & (A.7) \\ \overline{\theta_{0\dot{1}}} = c\theta_{0\dot{0}} + d\theta_{1\dot{0}}. & (A.8)\end{array}$$

The comparison of the eqs. $(A.5)$ and $(A.7)$ yields immediately $b = 0$ and $d = \bar{a}$ and since also $ad - bs = 1$ one has $a = \exp(ir_1), d = \exp(-ir_1)$ for some real $r_1$. Thus

$$\theta_{0\dot{0}} = e^{-ir_1/2}\Theta^1 \qquad (A.9)$$

where $\Theta^1$ is nonzero real form and besides

$$\theta_{0\dot{1}} = c\theta_{0\dot{0}} + e^{-ir_1}\theta_{1\dot{0}}.$$

Now let us apply the rotation $g_0\left(c\exp\left(-ir_1\right)\right)$. It does not affect all the dotted $S$-forms including $S_{\dot{0}}$. Denoting transformed tetrad as $g\theta$ one has as before

$$\overline{g\theta_{0\dot{0}}} = \overline{\theta_{0\dot{0}}} = e^{ir_1}g\theta_{0\dot{0}}$$

but

$$\overline{g\theta_{0\dot{1}}} = \overline{\theta_{0\dot{1}}} = e^{-ir_1}g\theta_{1\dot{0}}.$$

Hence there exists 1-form $\Theta^2$ such that

$$g\theta_{0\dot{1}} = e^{ir_1/2}\Theta^2, \quad g\theta_{1\dot{0}} = e^{-ir_1/2}\overline{\Theta^2}.$$

It is clear that $\Theta^2$ is not real and moreover the sequence of 1-forms $\Theta^1, \Theta^2, \overline{\Theta^2}$ is linearly independent in total.

Let us consider now $gS_2 = g\theta_{1\dot{0}} \wedge g\theta_{1\dot{1}} = e^{-ir_1/2}\overline{\Theta^2} \wedge g\theta_{1\dot{1}}$. Conjugated 2-form $\overline{gS_2}$ will be wedge orthogonal to 2-form $2gS_1 = g\theta_{0\dot{1}} \wedge g\theta_{1\dot{0}} + g\theta_{0\dot{0}} \wedge g\theta_{1\dot{1}} = -\Theta^2 \wedge \overline{\Theta^2} + e^{-ir_1/2}\Theta^1 \wedge g\theta_{1\dot{1}}$ that yields the equations

$$\left.\begin{array}{c}\Theta^2\\\overline{\Theta^2}\end{array}\right\} \wedge \Theta^1 \wedge g\theta_{1\dot{1}} \wedge \overline{g\theta_{1\dot{1}}} = 0.$$

They imply

$$\overline{g\theta_{1\dot{1}}} = k\,g\theta_{1\dot{1}} + l\Theta^1 \quad \Rightarrow \quad \overline{gS_2} = e^{-ir_1/2}\overline{\Theta^2} \wedge (k\,g\theta_{1\dot{1}} + l\Theta^1)$$

for some $k, l$. But it follows from equation $gS_2 \wedge \overline{gS_2} = 0$ and linear independence of $\Theta^1, \Theta^2, \overline{\Theta^2}, g\theta_{1\dot{1}}$ that $l = 0$ and hence $g\theta_{1\dot{1}} = \bar{k}\,\overline{g\theta_{1\dot{1}}} = |k|^2 g\theta_{1\dot{1}}$ yielding $k = e^{-ir_2}$ for some real $r_2$.

We infer an existence of real 1-form $\Theta^3$ such that

$$g\theta_{1\dot{1}} = e^{-ir_2/2}\Theta^3.$$

The forms collection $\{\Theta^1, \Theta^2, \overline{\Theta^2}, \Theta^3\}$ constitutes the basis of complexified covector space $\Lambda$.

Let us calculate finally the complex conjugate to

$$2gS_1 = -\Theta^2 \wedge \overline{\Theta^2} + e^{-i(r_1+r_2)}\Theta^1 \wedge \Theta^3.$$

Since $\Theta^1, \Theta^3, r_1, r_2$ are real the equation $gS_1 \wedge \overline{gS_1} = 0$ yields immediately $r_1, r_2 = 0 \ (mog\ 2\pi)$. Hence $g\theta_{1\dot{1}} = \varepsilon e^{ir_1/2}\Theta^3$ where $\varepsilon = 1$ or $\varepsilon = -1$. Let us denote $\hat{\Theta}^3 = \varepsilon \Theta^3$ and write out together the relations obtained above

$$g\theta_{0\dot{0}} = e^{-ir_1/2}\Theta^1, \quad g\theta_{0\dot{1}} = e^{ir_1/2}\Theta^2, \quad g\theta_{1\dot{0}} = e^{ir_2/2}\overline{\Theta^2}, \quad g\theta_{1\dot{1}} = e^{ir_1/2}\hat{\Theta}^3.$$

One sees that the metric tensor
$$\mathbf{g} = -g\theta_{A\dot{B}} \otimes g\theta^{A\dot{B}} = 2\,\Theta^2 \otimes_S \overline{\Theta^2} - 2\,\Theta^1 \otimes_S \hat{\Theta}^3$$
that corresponds to the transection is real and has Lorentz signature. The real pseudo-orthonormal basis of the covector space is formed by the sequence $\Theta^1 + \hat{\Theta}^3, \Theta^1 - \hat{\Theta}^3, \Theta^2 + \overline{\Theta^2}, i\Theta^2 - i\overline{\Theta^2}$. The theorem is proven. \hfill Q.E.D.